# Multi-path multi-component self-accelerating beams through spectrum-engineered position mapping

Yi Hu[1,2], Domenico Bongiovanni[1], Zhigang Chen[2,3], Roberto Morandotti[1]

[1]INRS-EMT, 1650 Blvd. Lionel-Boulet, Varennes, Québec J3X 1S2, Canada

[2]*The MOE Key Laboratory of Weak Light Nonlinear Photonics, School of Physics and TEDA Applied Physics School, Nankai University, Tianjin 300457, China*

[3]*Department of Physics and Astronomy, San Francisco State University, San Francisco, California 94132, USA*

*Corresponding author: morandotti@emt.inrs.ca*

We introduce the concept of spatial spectral phase gradient, and demonstrate, both theoretically and experimentally, how this concept could be employed for generating single- and multi-path self-accelerating beams. In particular, we show that the trajectories of the accelerating beams are determined *a priori* by different key spatial frequencies through direct spectrum-to-distance mapping. In the non-paraxial regime, our results clearly illustrate the breakup of Airy beams from a different perspective, and demonstrate how circular, elliptic or hyperbolic accelerating beams can be created by judiciously engineering the spectral phase. Furthermore, we found that the accelerating beams still follow the predicted trajectory also for vectorial wavefronts. Our approach not only generalizes the idea of Fourier-space beam engineering along arbitrary convex trajectories, but also offers new possibilities for beam/pulse manipulation not achievable through standard direct real-space approaches or by way of time-domain phase modulation.

Self-accelerating beams, featured by a transversely bending trajectory, have attracted a great deal of attention. Such an interest was initially stimulated by the intriguing properties of Airy waves, first introduced in quantum mechanics [1], and recently in optics [2, 3]. Indeed, self-accelerating Airy waves have been demonstrated and proposed for a host of potential applications in optics and a variety of other settings, including for example optical trapping and manipulation [4], plasma guidance [5], generation of light bullets [6], and routing of electrons [7]. An ideal Airy beam propagates non-diffractively with a small bending angle in the paraxial limit, but it tends to break up when bending into a large angle. Furthermore, non-paraxial self-accelerating beams that can bend along circular trajectories were predicted and demonstrated very recently [8-10]. Thus far, much of the progress in harnessing self-accelerating beams relied on specific monotonic phase modulations imposed in either real or Fourier space, leading to single-path propagation only [11-15]. It is then natural to ask: what would happen if a non-monotonic phase were introduced? Is it then possible to develop a generalized approach through spectral phase engineering that could give rise to large-angle self-accelerating beams with arbitrary trajectories? Although the angle of a ray is related to the spatial frequency, such an intuitive picture requires fundamental understanding as well as further experimental ground in order to properly explain the properties of self-accelerating beams.

In our recent work, we showed that it is possible to map the spectrum into the propagation distance via the use of suitable nonlinearities [16], suggesting a possible direct mapping between the spectrum and the path of self-accelerating beams. Such a feature would be even more desirable if it could be applied in the linear regime. By noticing that the spectra of all self-accelerating beams reported so far are composed of different spatial frequencies whose phases have a specific relationship, we are readily reminded of non-stationary signals in the time domain. For example in signal processing, the gradient spectral phase has been successfully introduced to describe the group delay [17]. Since the concept of group delay (indeed related to time) can be analogous to that of position in the spatial domain, it is then possible to perform the direct mapping from the spectrum to the beam path by introducing the concept of spatial spectral phase gradient. Such a linear spectrum-engineered position mapping has certainly several benefits towards the understanding and control of self-accelerating beams.

In this Letter, we propose and demonstrate a versatile approach to generate single- or multi-path self-accelerating beams through direct spectrum-to-distance mapping. We found that the path of a self-accelerating beam at different propagation distances is mediated by different key spatial frequencies, as confirmed by our direct experimental observation. Such a principle is the origin of many properties associated with accelerating beams, including self-bending and self-healing. Multiple trajectories (each of them being a convex curve) of self-accelerating beams are established with proper spectral phase and are managed by different parts of the spectrum. Under the non-paraxial condition, unlike circular, elliptical and hyperbolic trajectories that persist at large bending angles, our scheme reveals clearly that the cubic phase employed for the paraxial Airy beams leads to three interfering trajectories that break down the Airy beams. In essence, our scheme not only can trace the beam path for a known phase mask, but also permits imposing virtually any spectral phase configuration to achieve any desired convex trajectory. Furthermore, we demonstrate that the generalized method is also applicable for analyzing vector self-accelerating beams, where the intensity patterns including two polarization components are found to follow the trajectories predicted by our model.

As a proof of principle for our approach, let us consider a simple and realistic scenario: we modulate (via phase) a beam in the $x'$-$y'$ (i.e. the Fourier) plane [Fig. 1(a)], rather than in the real space [11, 12], to generate a self-accelerating beam. In the one-dimensional (1D) case, a cylindrical lens (with focal length $f$) and a $y'$-independent phase modulation are employed. By setting the polarization of the initial beam (wave number is $k$) along the $y'$ axis, the spectral evolution under the *paraxial condition* is described by the simple expression:

$$\tilde{E}(k_x, z) = \exp[i\mu(k_x, z)], \tag{1}$$

where $\mu(k_x, z) = -k_x^2 z/(2k) + \rho(k_x)$, $k_x = x'k/f$ is the spatial angular frequency and $\rho(k_x)$ is the imposed phase in the $x'$-$y'$ plane. As mentioned before, while in time domain the spectral phase gradient is related to the group delay [17], in the spatial domain the same quantity is associated with the local position of the light beam, i.e.,

$$x = -\partial \mu(k_x, z)/\partial k_x = k_x z/k - \rho'(k_x). \tag{2}$$

In a finite spatial regime defined by $\Delta x$, the beam is mainly composed (in the spectral picture) by the frequency components $\Delta k_x \partial^2 \mu(k_x, z)/\partial k_x^2$. Since the weight of each $k_x$ component is identical [see Eq. (1)], the spectral density ($\Delta k_x/\Delta x$) within $\Delta x$ is $1/|\partial^2 \mu(k_x, z)/\partial k_x^2|$. Thus the beam can reach a "spectral density singularity" (which is indeed related to the trajectory of self-accelerating beams) when the condition $\partial^2 \mu(k_x, z)/\partial k_x^2 = 0$ is satisfied, i.e.,

$$\rho''(k_x) = z/k. \tag{3}$$

By solving the above equation, the key spatial frequencies responsible for the beam paths (denoted by $k_{xc}$) can be obtained as a function of $z$. Let us first discuss the simple case in which we assume $\rho''(k_x)$ to be (smoothly) monotonically dependent on the frequency. Under this assumption, the key frequency determined by Eq. (3) must be single-valued along the propagation, indicating a single path of a self-accelerating beam. Using this finding together with Eq. (2), one can readily obtain that the trajectory of the beam follows a convex curve. A typical example for the previous discussion is shown in Figs. 1(b) and (c) by analyzing a cubic phase $\rho(k_x) = -(\alpha k_x / k)^3$ ($\alpha$=200) related to the well-known Airy beam case [2]. Replacing $k_x$ in Eq. (2) with the key frequency $k_{xc} = -k^2 z / (6\alpha^3)$ [plotted in Fig. 1(b) and obtained from Eq. (3)], the trajectory is expected to follow a parabolic path $-kz^2/(12\alpha^3)$, as further confirmed through the beam evolution calculated by numerically Fourier transforming Eq. (1) [Fig. 1(c)]. If we discard all the sub-lobes of the beam, the spatial frequency of the residual main lobe obviously evolves as the key frequency during propagation [Fig. 1(b)]. More precisely, as a result of the additional influence of the spectral components far from the key frequency, the location of the intensity maximum (IM) tends to slightly deviate from the one relating to the "spectral density singularity". For instance [see the inset in Fig. 1(c)], the "spectral density singularity" is estimated to appear at $x$=0, which almost matches the IM calculated from a small spectral segment around the key frequency (upper panel), but has a slight dislocation with respect to the IM corresponding to a large enough frequency scale (bottom panel). In spite of the slight mismatch, the trajectory calculated by Eq. (2) and (3) perfectly predicts the path of the main hump [Fig. 1(c)]. Coming back to the general case of a single beam path, illustrated by the example of a cubic phase applied, the main hump at different propagation distances is managed by different key spatial frequencies. This "frequency-uncorrelated" propagation of the main hump is the origin of the self-healing properties [18] typically associated to self-accelerating beams, indicating that even if the main lobe is filtered at a certain position, it tends to regenerate along propagation, so that its profile, particularly far away from the blocker, is not influenced.

Next, we consider a more general case in which $\rho''(k_x)$ is not necessarily a monotonic function. Thus two or more key frequencies tend to be involved, leading to multiple beam paths. As a typical example, the sinusoidal phase $\rho(k_x) = 130\sin(80 k_x / k)$ is analyzed and the results are shown in Figs. 1(d) and (e). As expected, the key frequencies are now multi-valued [see the curved lines in Fig. 1(d)]. Accordingly, three trajectories are calculated by inserting the key frequencies into Eq. (2) and each of them is managed by different parts of the spectrum (the same number in the notation is used for their correspondence in Figs. 1(d) and 1(e)). They predict the beam path obtained by numerically Fourier transforming Eq. (1). Keeping the main lobe of the $2^{nd}$ branch in Fig. 1(e), the residual spectral power follows the corresponding key frequency [Fig. 1(d)]. Such a correspondence is also observed (not shown here) for the other two beam paths. Although different parts of the frequency are responsible for different paths of the multi-trajectories, the key frequency is still monotonic in any of the single trajectories. Therefore, each beam path manifests the same features just as a single-path beam does.

The above analysis implies that the beam paths are predictable for a known spectral phase through employing the theory formulated via Eq. (2) and (3). To verify this, we perform the experiment in a setup similar to Fig. 1(a). A phase-only spatial light modulator (SLM) is placed at the front focal plane of a cylindrical lens ($f$=100 mm) in order to modulate a broad beam (featured by a wavelength of 633 nm). A camera positioned behind the lens is used to record both the beam evolution and its spatial spectrum (obtained by adding an additional cylindrical lens). Firstly, a properly engineered cubic phase [19], as shown in Fig. 2(a), is imposed on the beam. As it is well known, the resulting Airy beam propagates along a parabolic trajectory [Fig. 2(b)]. Akin to the simulation in Fig. 1(b), the corresponding spectrum for the filtered main hump shifts linearly with the propagation distance [Fig. 2(c)]. This is a direct evidence that the main hump of the self-accelerating beam corresponds to different spatial frequencies at different propagation distances. In the linear case, such a physical pattern can only be discovered through getting rid of the sub-lobes of the beam. However, in a nonlinear environment, it is directly revealed by the shifting of certain reshaped spectral defects [16]. Secondly, we employed a sinusoidal phase function, plotted in Fig. 2(d), to generate multi-path trajectories featured by three main humps. The measured beam propagation agrees very well with our predictions [Fig. 2(e)]. For example when we alternatively block two beam paths in the real space, we find that each of the remaining beam paths is managed by different parts of the spectrum [Fig. 2(f)], as expected by the analysis in Fig. 1. Although the examples presented here are for trajectory prediction from a given phase function, we can also calculate the phase structure related to any desired (convex) trajectory.

The above analysis is also applicable to the *non-paraxial condition*, where the spectrum evolution of a beam takes the following form: $\tilde{E}_{np}(k_x,z) = \exp[i(k^2 - k_x^2)^{1/2}z + i\rho(k_x)]$. Using a similar procedure as that used in Eq. (1), the beam trajectories are also predictable for a given phase $\rho(k_x)$ and they manifest the same behaviors as in the paraxial case: different trajectories correspond to different parts of the spectrum and the key frequency along a single trajectory monotonically varies during propagation. These features restrict all the beam paths to follow convex trajectories. Two typical examples are shown in Figs. 3(a) and 3 (b). One corresponds to the inverse sinusoidal function $\rho(k_x) = r\sin^{-1}(k_x/k)$ used to generate Bessel-like self-accelerating beams [10]. Using our method, we can still predict that the beam will follow a circular trajectory $T_r(z) = -(r^2/k^2 - z^2)^{1/2}$, as shown in Fig. 3(a). Differently from the work in Ref. [10], no spectral amplitude modulation is employed here. The other example is associated with a cubic phase $(5k_x/k)^3$. As we know, in the paraxial approximation, this phase corresponds to an Airy beam [2]. However, in the non-paraxial case, such a phase structure leads to a beam showing three main trajectories, whose interference induces a strong deformation of the Airy beam [Fig. 3(b)]. This basically explains why Airy beams cannot persist under the non-paraxial conditions from a perspective not apparent in all previous studies. We can see here that, in the non-paraxial regime, a monotonic phase is also possibly related to multi-trajectories.

Conversely, beam paths can be controlled by producing a proper phase modulation. As a typical example, Figure 3(c) shows the numerically-obtained phase shape relative to elliptical and hyperbolic trajectories, i.e., $a(b^2 - z^2)^{1/2}/b$ and $c(d^2 + z^2)^{1/2}/d$, respectively. All the frequencies participate in forming elliptical self-accelerating beams while only a part of the frequencies ($k_x/k \leq |c/(c^2+d^2)^{1/2}|$) is responsible for the beam trajectory of hyperbolic self-accelerating beams [Fig. 3(c)]. Typically, the values $a$=10μm, $b$=30μm, $c$=$d$=30μm are employed. By imposing the estimated phases in the frequency domain, the main hump of the beams follows the desired trajectories, as shown in Figs. 3(d) and (e), respectively. Compared with the method used in [14, 15] where the phase required for the beam to follow an elliptical path is obtained by solving the complex angular Mathieu functions, the approach shown here is much easier and more straightforward in manipulating the trajectories of self-accelerating beams. For multi-path self-accelerating beams along two or more different trajectories, the procedure is similar, provided that the spectral portions associated to different trajectories are non-overlapping.

In the previous analysis, we only discussed the case of scalar (corresponding to a $y'$-polarized input) self-accelerating beams. However, if the initial input depicted in Fig. 1(a) is $x'$-polarized, the polarization should be taken into account through the use of Jones matrix [20]. Then the spectral evolution of a beam is composed of both $x$- and $z$- components: $\tilde{E}_x = k^{-1}(k^2 - k_x^2)^{1/2}\tilde{E}_{np}$, $\tilde{E}_z = -k^{-1}k_x\tilde{E}_{np}$. Although additional amplitude modulations are included in the above equations, the previous method (used to analyze the scalar non-paraxial case) is still applicable, as the spatial spectra in $\tilde{E}_x$ and $\tilde{E}_z$ are complementary. Figure 4 shows a typical example associated with a sinusoidal phase $\rho(k_x)$=100sin(2π$k_x$/$k$). The key frequencies for both components are solved by only considering the phase term [Fig. 4(a)]. As expected, the calculated trajectories have a good matching with the beam evolution of the total intensity [Fig. 4(b)]. One can clearly tell the complementarity in terms of fields of the *x*- and *z*-components by comparing Figs. 4(c) and (d) with their sum in Fig. 4(b). These results indicate that our method is still applicable for vector self-accelerating beams. Straightforwardly, it can be envisaged that this scheme is useful for analyzing the 3D case under the non-paraxial condition, where the polarization issue always needs to be considered [15].

In summary, we demonstrated a generalized scheme to generate and control single- or multi-path self-accelerating beams through spectrum-engineered position mapping in both paraxial and non-paraxial conditions. In analogy to what happens in the time domain, the gradient of the spatial spectral phase is related to the beam location. In this new framework, we found that different trajectories are associated to different parts of the spectrum, and in a single trajectory, the key frequency for the main hump of the beam monotonously varies along propagation. The properties of self-accelerating beams can be now understood in terms of frequency. In the non-paraxial condition, we clearly illustrate the reason for the breakup of Airy beams, whereas any large-angle self-accelerating beam along a convex trajectory can be generated through our method. What we believe is even more important is that our scheme can also be applied to vector self-accelerating beams. There are many fundamental issues that merit future investigations, such as how to extend our method to incoherent light, how to manipulate spatiotemporal self-accelerating beams [6], and how to extend our scheme to the nonlinear regime [16]. In turn, such interesting questions can be easily generalized and lead to a better understanding of other self-accelerating waves present in nature.

This work is supported by the NSERC (Natural Sciences and Engineering Research Council of Canada) Discovery and Strategic grant programs, by US National Science Foundation and AFOSR, and by the State Key Program for Basic Research of China (No. 2013CB632703). Y. H.

acknowledges support from MELS FQRNT Fellowship.

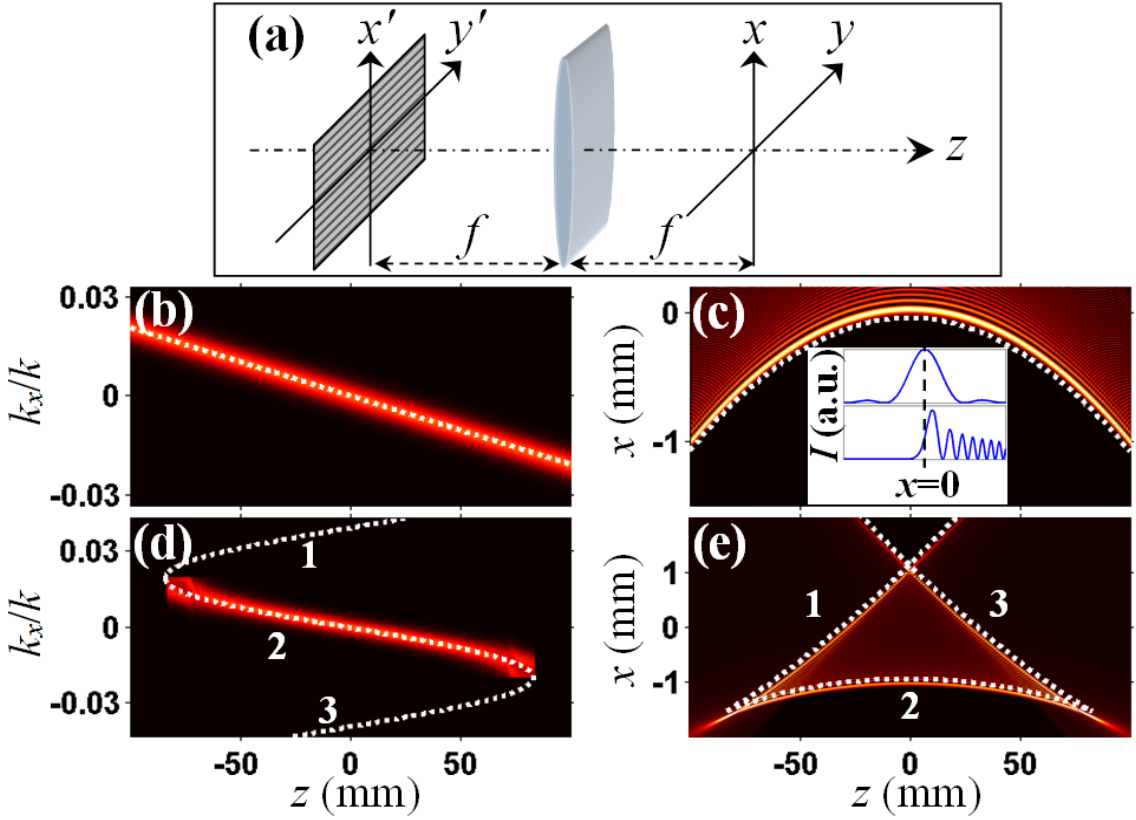

FIG. 1 (color online). (a) Sketch of experimental geometry used for the generation of self-accelerating beams. (b-e) Calculated spectrum (b,d) and beam propagation (c,e) with two typical examples of imposed cubic (2nd row) and sinusoidal (3rd row) phase under the paraxial approximation. White lines in (b, d) mark the key frequency corresponding to the main lobe of the accelerating beams in (c, e), while the predicted beam trajectories are traced by the dashed white curves in (c, e). Inserts in (c) show the transverse intensity profiles at $z=0$ as calculated from the frequency range under the constrains $|k_x|\leq 0.002k$ (upper panel) and $|k_x|\leq 0.043k$ (bottom panel). The numbers in (d, e) indicate the correspondence between the key spectral ranges and the resulting trajectories. (All calculated trajectories are slightly shifted with respect to the main lobes associated to beam evolution).

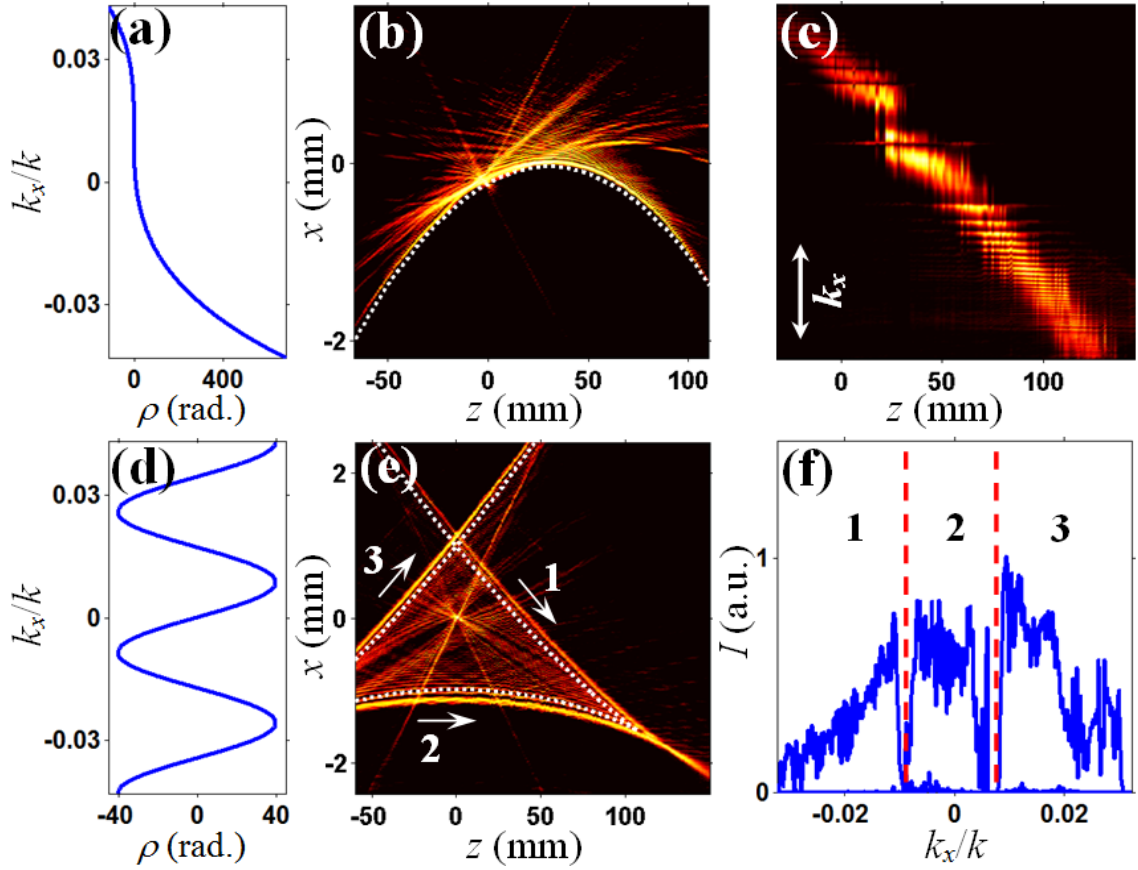

FIG. 2 (color online). Experimental results of self-accelerating beams obtained with a cubic (top panels) and a sinusoidal (bottom panels) phase similar to that presented in Fig. 1. (a,d) Spectral phase distribution; (b), (e) Single and triple trajectories of the accelerating beams resulting from (a) and (d), respectively; (c) the residual spectrum of the filtered main hump in (b); (f) Different spectral ranges responsible for different trajectories in (e).

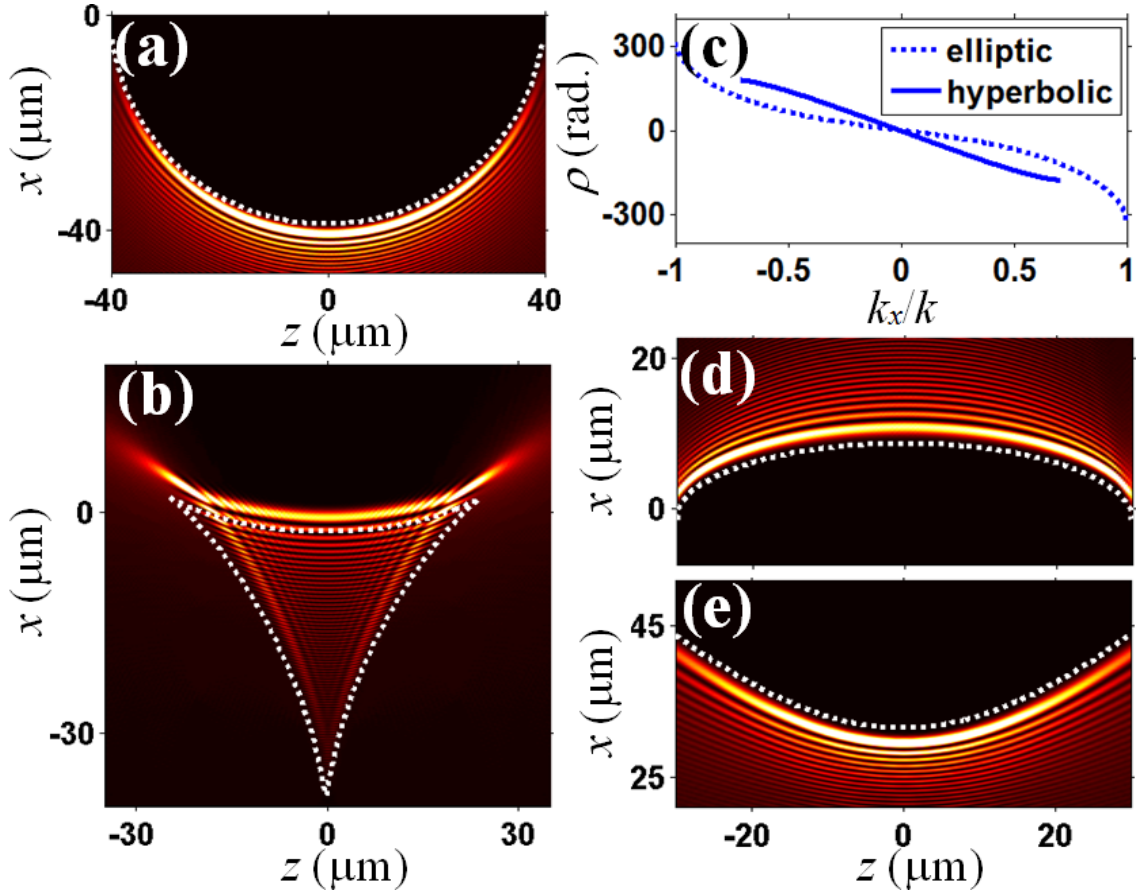

FIG. 3 (color online). Self-accelerating beams resulting from direct spectrum-to-distance mapping under the non-paraxial condition. (a) and (b) show the half-Bessel and Airy-like beam trajectories obtained by applying an inverse sinusoidal and cubic phase, respectively; (d) and (e) show the elliptic and hyperbolic beam trajectories obtained with the spectral phases plotted in (c).

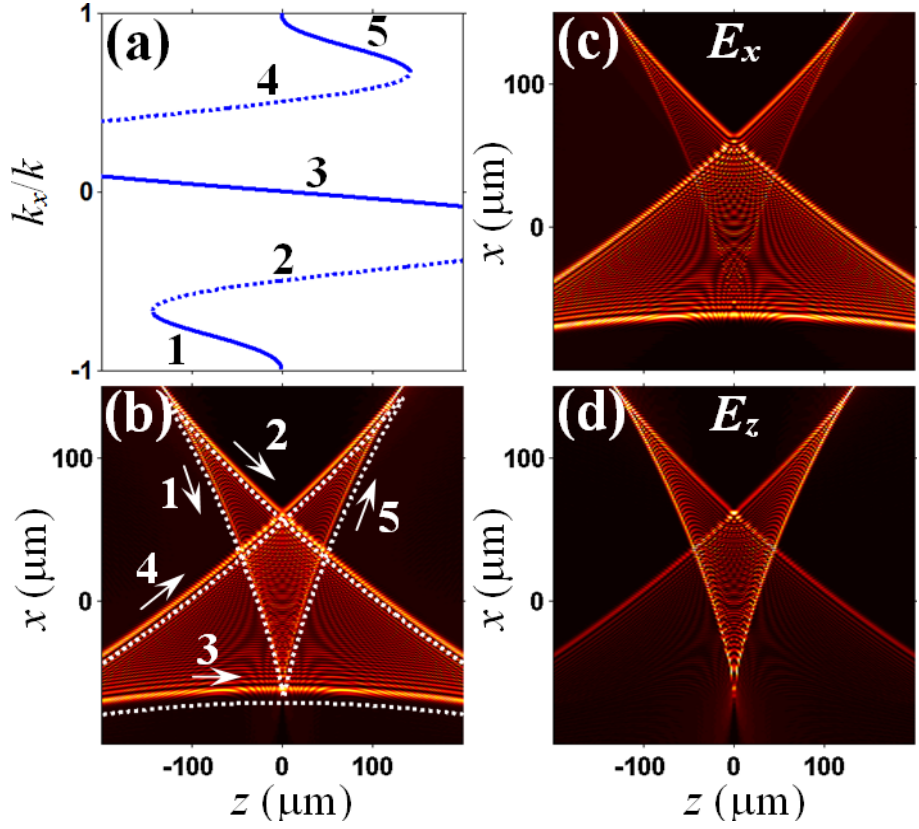

FIG. 4 (color online). Vector self-accelerating beams obtained under the non-paraxial condition. (a) shows the key frequency distribution, (b) the total intensity of the beam, (c) and (d) the intensity patterns for the $x$- and $z$-components, respectively. The numbers in (a, b) indicate the correspondence between the key spectral ranges and the resulting trajectories. Note that (b) is given by the sum of (c) and (d).